\documentclass{ws-procs975x65}
\usepackage{mathrsfs}
\usepackage{graphicx}
\def\beq{\begin{equation}}
\def\eeq{\end{equation}}

\newcommand{\scrim}{{$\mathscr{I}^-$}}
\newcommand{\hp}{{$\mathscr{H}^+$}}
\newcommand{\hm}{{$\mathscr{H}^-$}}
\begin{document}

\title{Do not mess with time:\\Probing faster than light travel and chronology protection\\ with superluminal warp drives}
\author{Stefano Liberati$^*$}

\address{SISSA,
Via Bonomea 265, 34136 Trieste, Italy\\
and INFN, sez. Trieste.\\
$^*$E-mail: liberati@sissa.it\\
http://people.sissa.it/$\sim$liberati}

%\author{Anthony N. Author} 
%
%\address{Group, Laboratory, Street,\\
%City, State ZIP/Zone, Country\\
%E-mail: an\_author@laboratory.com}

\begin{abstract}
While General Relativity (GR) ranks undoubtedly among the best physics theories ever developed, it is also among those with the most striking implications. In particular, GR admits solutions which allow faster than light motion and consequently time travel. Here we shall consider a ``pre-emptive" chronology protection mechanism that destabilises superluminal warp drives via quantum matter back-reaction and hence forbids even the conceptual possibility to use these solutions for building a time machine. This result will be considered both in standard quantum field theory in curved spacetime as well as in the case of a quantum field theory with Lorentz invariance breakdown at high energies. Some lessons and future perspectives will be finally discuss.
\end{abstract}


\bodymatter

%%%%%%%%%%%%%%%%% now a standard article style for the most part

\section{Introduction}

After one hundred year GR still stands strong having passed many observational tests and it is nowadays applied in everyday life (e.g.~in GPS based devices). Nonetheless, there are still many puzzling predictions of GR which seems to stretch to the limit our notion of reality. Among these one can obviously lists singularities and holography as well as solutions which allow superluminal and time travels. 

\section{Time travel in a nut-shell}

Technically time travel is associated to the notion of Closed Timelike Curves (CTC), i.e.~to the possibility for physical observers to move on close paths. 
%It is evident that such possibility is ``per se" problematic as it implies that ones' journey can end up to where it started in space {\em and} time. 
Once this possibility is realised there is no obstruction for an observer to appear in the same space position even before it started its journey, i.e. to travel backward in time.%\footnote{Of course one can conceive also travelling forward in time, but this occurrence has no real paradoxes associated and indeed it is already possible within special and general relativity (e.g.~the ``twins' paradox" (which is not a real logical paradox) of special relativity, or the effects of gravitational time delation in GR.}

Real time machines are fun and useful for the entertainment industry but are very pernicious from a physicists' point of view. The basic point being that they immediately lead to logical paradoxes. These are often divided in two large families: the so called ``grand father paradoxes" and the ``bootstrap paradoxes".

The first kind of paradox is referring to the logical inconsistency which can be generated if someone travelling back in time would change events whose future developments led, more or less directly, to his/her very existence or time travel. For example one can think of the paradox associated to the fact that by travelling back in time we could kill one of our ancestors and so in principle prevent our very possibility to come into existence in first place.

The second kind is slightly more subtle but still quite puzzling. Indeed, backwards time travel would allow for causal loops involving events or informations whose histories form a closed loop, and thus seem to ``come from nowhere". For example, think of the causal loops where at some point ones travels back in time to give to his/her past-self the numbers associated to some lottery. 
%By doing this one's life would follow from that moment on a completely different path but at some point logical consistency would require he/she to travel back in time again to give those numbers again to his/her past self as if he/she would decide or forget to do so it would be a paradox that he/she got the numbers in the first place.

%Several solutions have been proposed for such paradoxes. Let us briefly review some of them~\cite{Visser:1995cc}.
%{Branching universe}: the universe branches continuously and if one travels back in time changing history it would simply create a new branch (or move to a branch different the time traveller came from in the first place). This option would probably imply a radical rewriting of physics law implementing them on what are technically called non-Hausdorff manifolds.
%{Novikov's conjecture}: the logical consistency requirement implies that only periodic solutions are allowed. If one travels back to the past and tries to change history not only he/she will be doomed to fail but his/her action would most probably implement things that had happened so that the history went the way it did.
%{Hawking's Chronology Protection conjecture}: time machines are forbidden by physical laws, in particular quantum effects on curved spacetime. So time machines must be intrinsically unstable and if one tries to form one he/she will necessary fail.   
%
Several solutions have been proposed for such paradoxes (see e.g.~for a detailed discussion Ref.~\refcite{Visser:1995cc}) we shall here focus on the so called
{Hawking's chronology protection conjecture} which states that time machines are forbidden by physical laws, in particular quantum effects on curved spacetime. So time machines must be intrinsically unstable and any attempt to establish one will necessary fail. Note that by generation of a time machine we mean here that a time orientable spacetime (with a definitive time orientation) endowed with a Òcausally innocuous pastÓ (i.e.~with no CTC) presents in the future some CTCs. The boundary of the region characterised by the presence of CTCs is a Cauchy horizon also called the chronology horizon. 

%
%\begin{itemize}
%\item{Branching universe}: the universe branches continuously and if one travels back in time changing history it would simply create a new branch (or move to a branch different the time traveller came from in the first place). This option would probably imply a radical rewriting of physics law implementing them on what are technically called non-Hausdorff manifolds.
%\item{Novikov's conjecture}: the logical consistency requirement implies that only periodic solutions are allowed. If one travels back to the past and tries to change history not only he/she will be doomed to fail but his/her action would most probably implement things that had happened so that the history went the way it did.
%\item{Hawking's Chronology Protection conjecture}: time machines are forbidden by physical laws, in particular quantum effects on curved spacetime. So time machines must be intrinsically unstable and if one tries to form one it will necessary fail.   
%\end{itemize}

%In this brief report we shall focus on the last conjecture. 
The chronology protection conjecture had in recent years a growing support by several explorative results with quantum field theories in curved spacetime.~\cite{Visser:2002ua} Unfortunately, a well known theorem by Kay, Radzikowski, and Wald implies that on chronological horizons there are always points where the two-point function is not of the Hadamard form which in turns implies that at these points the entire process of defining a renormalized stress-energy tensor breaks down.\cite{Kay:1996hj} 

However, a renormalized stress energy tensor is what one needs in order to properly describe the back-reaction of the quantum effects on the geometry vie the semiclassical Einstein equations. Henceforth, all hopes to settle down the validity of Hawking's conjecture within the realm of standard physics seem to be lost. Indeed, nowadays the most common opinion is that the chronology protection conjecture will have to wait for a full fledged quantum gravity theory in order to be settle.\cite{Visser:2002ua}

In what follows, we shall offer a glimpse of hope, showing that at least in some cases a sort of ``pre-emptive" chronology protection holds, in the sense that those spacetimes which could be used in order to generate a time machine might turn out to be  unstable to quantum effects and hence not even in principle useful for attempting the construction of a time machine. 
%In order to do this we need however to spend a word first on what are these dangerous spacetimes.

\section{Time machines spacetimes and warp drives}

There are several ways in which time machines arise in GR.
In brief, there are basically two families of solutions: rotating spacetimes and spacetimes which allow for faster than light travel.
Among rotation induced time machines one can list, G\"odel Universe, Tipler's and Gott's Time machines (rotating dust cylinders or strings), and the very same Kerr black hole (see Ref.~\refcite{Visser:1995cc} and references therein).
Apart from the last one these are generally seen more as an evidence that GR per se is not protected from CTC if sufficiently contrived set-ups are conceived, while in the case of the Kerr solution the CTC are confined behind the inner (Cauchy) horizon, where the solution is anyway not considered trustworthy anymore.

Spacetime solutions allowing superluminal travel are much harder to dismiss. Warp drives and traversable wormholes are very popular in science fiction but are also honest solutions of GR which apart from requiring substantial quantities of exotic (specifically null energy conditions violating) matter, do not seem to have {\em per se} inconsistencies or classical instabilities.

The warp-drive geometry was introduced by Miguel Alcubierre in 1994 (see Ref.~\refcite{Alcubierre:1994tu}) and represents a bubble containing an almost flat region, moving at arbitrary speed within an asymptotically flat spacetime. Mathematically its metric can be written as
\begin{equation}\label{eq:3Dalcubierre}
ds^2=-c^2 dt^2+\left[dx-v(r)dt\right]^2+dy^2+dz^2~,
\end{equation}
where $r\equiv \sqrt{[x-x_c(t)]^2+y^2+z^2}$ is the distance from the center of the bubble, $\{x_c(t),0,0\}$, which is moving in the $x$ direction with arbitrary speed $v_c=dx_c/dt$. Here $v(r)=v_c f(r)$ and $f$ is a suitable smooth function satisfying $f(0)=1$ and $f(r) \to 0$ for $r\to \infty$.  
%In Fig.~\ref{fig:wd-stru}, the curvature of the warp-drive geometry is plotted: 
To make the warp-drive travel at the speed $v_c(t)$, the spacetime has to contract in front of the warp-drive bubble and expand behind it. It is easy to see that the worldline $\{x_c(t),0,0\}$ is a geodesic for the above metric. Roughly speaking, if one places a spaceship at $\{x_c(t),0,0\}$, it is not subject to any acceleration, while moving faster than light with respect to someone living outside of the bubble (here the spaceship is basically treated as a test particle, see~Ref.~\refcite{Lobo:2004wq} for a more general treatment). 
%The spaceship inside the warp drive bubble is as a matter of fact isolated from the spacetime surroundings and cannot interact with it, however one could in principle conceive to built a sort of ``interstellar railway" running from Earth to a distant planet which by a coordinated generation of energy violating matter could locally produce and move a warp drive, with a spaceship inside, at superluminal speeds.

Of course this sounds pretty exciting but on second thoughts also very worrisome. Indeed, it is quite easy to transform any superluminal travel capable structure into a time machine. In the case of the warp-drive this would only require a two way trip at arbitrary speeds as this is well known to be able to generate CTC.\cite{Liberati:2001sd}
%In conclusion, GR seems to predict the possibility of superluminal and time travel at the  moderate cost to solve the engineering problem to produce and control exotic matter.
However, as usual, ``the devil is in the details" as we shall see next.
 
%-------------------------------------------------------------
\section{Superluminal warp drive instability}
\label{sec:calculation}
%-------------------------------------------------------------

We are going to discuss now the instability associated with a superluminal warp drive. In the actual computation we shall restrict our attention to the $1+1$ dimensions case (since in this case one can carry out a complete analytic treatment).
%\footnote{However, we do expect that the salient features of our results would be maintained in a full 3+1 calculation, given that they will still be valid in a suitable open set of the horizons centred around the axis aligned with the direction of motion.}
Changing coordinates to those associated with an observer at the center of the bubble, the warp-drive metric~\eqref{eq:3Dalcubierre} becomes
\begin{equation}\label{eq:fluidmetric}
 ds^2=-c^2 dt^2+\left[dr-\bar{v}(r)dt\right]^2,\qquad \bar{v}=v-v_c~,
\end{equation}
where $r\equiv x-x_c(t)$ is now the signed distance from the center. Let us consider a dynamical situation in which the warp-drive geometry interpolates between an initial Minkowski spacetime [$\hat{v}(t,r)\to0$, for $t \to -\infty$] and a final stationary superluminal ($v_c>c$) bubble [$\hat{v}(t,r)\to\bar{v}(r)$, for $t \to +\infty$]. To an observer living inside the bubble this geometry has two horizons, a \emph{black horizon} \hp~located at some $r=r_1$ and a \emph{white horizon} \hm~located at $r=r_2$. 
%Fig.~\ref{fig:CP} show the Carter--Penrose diagram of these spacetimes. 
Here let us just add that from the point of view of the Cauchy development of \scrim~these spacetimes posses Cauchy horizons.
%
%\begin{figure}[h]
%\begin{center}
%\includegraphics[width=3in]{fig3}
%\end{center}
%\caption{Carter--Penrose diagram of the dynamical generation of a superluminal warp drive from flat spacetime.}
%\label{fig:CP}
%\end{figure}
%------------------------------------
\subsection{Light-ray propagation}
%------------------------------------

Let us now consider light-ray propagation in the above described geometry. Only the behaviour of right-going rays determines the universal features of the RSET, just like outgoing modes do in the case of a black hole collapse (see Ref.~\refcite{Finazzi:2009jb,Barcelo:2006np,Barcelo:2007yk}). Therefore, we need essentially the relation between the past and future null coordinates $U$ and $u$, labelling right-going light rays. 
%(see {  Fig.}~\ref{fig:lightrayssimp}). 
%Following \cite{particlecreation}, {  this relation can be found by integrating the {  right-going-ray} equation} 
%%
%\begin{equation}\label{eq:rays}
% \frac{dr}{dt}=c+\hat{v}(r,t)~.
%\end{equation}
%%
%
%{  
There are two special right-going rays defining, respectively, the asymptotic location of the black and white horizons. In terms of the right-going past null coordinate $U$ let us denote these two rays by $U_{\rm BH}$ and $U_{\rm WH}$, respectively. The finite interval $U \in (U_{\rm WH}, U_{\rm BH})$ is mapped to the infinite interval $u \in (-\infty, +\infty)$ covering all the rays traveling inside the bubble. 
For rays which are close to the black horizon, the relation between $U$ and $u$ can be approximated as a series of the form\cite{Finazzi:2009jb}
\begin{eqnarray}\label{eq:Uu}
U(u\to +\infty) \simeq U_{\rm BH} + A_1 e^{-\kappa_1 u} + \frac{A_2}{2}e^{-2\kappa_1 u} + \dots ~.
\end{eqnarray}
Here $A_n$ are constants (with $A_1<0$) and $\kappa_1>0$ represents the surface gravity of the black horizon. This relation is the standard result for the formation of a black hole through gravitational collapse. As a consequence, the quantum state which is vacuum on \scrim~will show, for an observer inside the warp-drive bubble, Hawking radiation with temperature $T_{H}=\kappa_1 /2\pi$.

Equivalently, we find that the corresponding expansion in proximity of the white horizon is~\cite{Finazzi:2009jb} 
\begin{eqnarray}
U(u\to -\infty) \simeq U_{\rm WH} + D_1 e^{\kappa_2 u} + \frac{D_2}{2}e^{2\kappa_2 u} + \dots ~,
\end{eqnarray}
where $D_2>0$ and $\kappa_2$ is the white hole surface gravity and is also defined to be positive. The interpretation of this relation in terms of particle production is not as clear as in the black horizon case and a full study of the renormalised stress energy tensor (RSET) is required.

%------------------------------------------------
\subsection{Renormalized stress-energy tensor}
%------------------------------------------------

In past null coordinates $U$ and $W$ the metric can be written as
\begin{equation}\label{eq:metricUW}
 ds^2=-C(U,W)dUdW~.
\end{equation}
In the {stationary region {at late times}, we can use the previous future null coordinate $u$} and a new coordinate $\tilde w$, defined as 
\begin{eqnarray}\label{eq:wdef}
\tilde w(t,r)= t + \int_{0}^{r} \frac{dr}{c-\bar{v}(r)}~.
\end{eqnarray}
In these coordinates the metric is expressed as 
\begin{equation}\label{eq:metricuw}
 ds^2=-\bar{C}(u,\tilde w) du d\tilde w~, \quad C(U,W) = \frac{\bar{C}(u,\tilde w)}{\dot{p}(u)\dot{q}(\tilde w)}~,
\end{equation}
where 
$U=p(u)$ and $W=q(\tilde w)$. In this way, $\bar{C}$ depends only on $r$ through $u,\tilde w$.

For concreteness, we refer to the RSET associated with a quantum massless scalar field living on the 
spacetime. The RSET components for this case are given in~Ref.~\refcite{Birrell:1982ix}
%\begin{equation}
%T_{UU} = -\frac{1}{12\pi}C^{1/2}\partial_U^2 C^{-1/2}~,\, T_{WW} = -\frac{1}{12\pi}C^{1/2}\partial_W^2 C^{-1/2}~,\, T_{UW} = T_{WU} =\frac{1}{96\pi}C~R~.\label{eq:TUWR}
%\end{equation}
%
%%
%\begin{align}
% T_{UU} &= -\frac{1}{12\pi}C^{1/2}\partial_U^2 C^{-1/2}~,\label{eq:TUU}\\
% T_{WW} &= -\frac{1}{12\pi}C^{1/2}\partial_W^2 C^{-1/2}~,\label{eq:TWW}\\
% T_{UW} &= T_{WU} =\frac{1}{96\pi}C~R~.\label{eq:TUWR}
%\end{align}
%%
%If there were other fields present in the theory, the previous expressions would be multiplied by a specific numerical factor. 
and using the relationships $U=p(u)$, $W=q(\tilde w$) and the time-independence of $u$ and $\tilde w$, one can calculate the RSET components in the stationary (late times) region.\cite{Finazzi:2009jb}
%\begin{align}
% T_{UU} &= -\frac{1}{48\pi}\frac{1}{\dot{p}^2}\left[\bar{v}'\,^2+\left(1-\bar{v}^2\right)\bar{v}\bar{v}''-f(u)\right]~,\\
% T_{WW} &= -\frac{1}{48\pi}\frac{1}{\dot{q}^2}\left[\bar{v}'\,^2+\left(1-\bar{v}^2\right)\bar{v}\bar{v}''-g(\tilde w)\right]~,\\
% T_{UW} &= T_{WU}=-\frac{1}{48\pi}\frac{1}{\dot{p}\dot{q}}\left(1-\bar{v}^2\right)\left[\bar{v}'\,^2+\bar{v}\bar{v}''\right]~,
%\end{align}
%where we have put $c=1$ and we have defined
%\begin{align}
% f(u)&\equiv\frac{3\ddot{p}^2(u)-2\dot{p}(u)\,\dddot{p}(u)}{\dot{p}^2(u)}~,\label{fu}\\
% g(\tilde w)&\equiv\frac{3\ddot{q}^2(\tilde w)-2\dot{q}(\tilde w)\,\dddot{q}(\tilde w)}{\dot{q}^2(\tilde w)}~.
%\end{align}
%One can show~\cite{wd} that $\dot{q}$ contains solely information associated with the dynamical details of the transition region. Moreover, for simple dynamical interpolations between Minkowski and the final warp drive, $\dot{q}(\tilde w )$ goes to a constant at late times, such that $g(\tilde w)\to 0$. From now on, we will neglect this term.

Let us now focus on the energy density inside the bubble, in particular at the energy $\rho$ as measured by a set of free-falling observers, whose four velocity is $u_{c}^\mu=(1,\bar{v})$ in $(t,r)$ components. For these observers neglecting transient terms one obtains~\cite{Finazzi:2009jb}
%\begin{equation}
$ \rho=T_{\mu\nu}u_{c}^\mu u_{c}^\nu=\rho_{\rm st}+\rho_{{\rm dyn}}$,
%~,\\
%\end{equation}
where we define a static term $\rho_{\rm st}$, depending only on the $r$ coordinate through $\bar{v}(r)$,
\begin{equation}
 \rho_{\rm st}\equiv-\frac{1}{24\pi}\left[\frac{\left(\bar{v}^4-\bar{v}^2+2\right)}{\left(1-\bar{v}^2\right)^2}\bar{v}'\,^2+\frac{2\bar{v}}{1-\bar{v}^2}\bar{v}''\right]~,\label{eq:rhocs}
 \end{equation}
  and a, time-dependent, dynamic term
  \begin{equation}
 \rho_{{\rm dyn}}\equiv\frac{1}{48\pi}\frac{{\cal F}(u)}{\left(1+\bar{v}\right)^2}~,\label{eq:rhodu} \quad\mbox{where}\quad 
{\cal F}(u)\equiv\frac{3\ddot{p}^2(u)-2\dot{p}(u)\,\dddot{p}(u)}{\dot{p}^2(u)}~.\\
% g(\tilde w)&\equiv\frac{3\ddot{q}^2(\tilde w)-2\dot{q}(\tilde w)\,\dddot{q}(\tilde w)}{\dot{q}^2(\tilde w)}~.
\end{equation}
%These latter term, depending also on $u$, corresponds to energy travelling on right-going rays, eventually red/blue-shifted by a term depending on $r$.

%------------------------------------------------
\subsection{Physical interpretation}
%------------------------------------------------

Let us start by looking at behavior of the RSET in the center of the bubble at late times. Here $\rho_{\rm st}=0$, because $\bar{v}(r=0)=\bar{v}'(r=0)=0$. 
%It is easy to see also that $u(t,r)$ is linear in $t$ so that, for fixed $r$, it acquires with time arbitrarily large positive values. 
One can evaluate $\rho_{\rm dyn}$ from Eq.~\eqref{eq:rhodu} by using a late-time expansion for ${\cal F}(u)$, which 
%(up to the first non-vanishing order in $e^{-\kappa_1 u}$) 
gives ${\cal F}(u) \approx \kappa_1^2$, so that $\rho(r=0) \approx \kappa_1^2/(48\pi)=\pi T_H^2/12$, where $T_H \equiv \kappa_1/(2\pi)$ is the usual Hawking temperature. 
%The above expression is the energy density of a scalar field in $1+1$ dimension at finite temperature $T_H$. 
This result confirms that an observer inside the bubble measures a thermal flux of radiation at temperature $T_H$.
%
%%----------------------------------------
%\subsection{Problems with horizons}
%%----------------------------------------

Let us now study $\rho$ on the horizons \hp~and \hm. Here, both $\rho_{\rm st}$ and $\rho_{{\rm dyn}}$ are divergent because of the $(1+\bar{v})$ factors in the denominators. Using the late time expansion of ${\cal F}(u)$ in the proximity of the black horizon~(see Ref.~\refcite{Finazzi:2009jb}) one gets
\begin{equation}
\lim_{r\to r_1 }{\cal F}(u) = \kappa_1^2\left\{1+\left[3{\left(\frac{A_{2}}{A_{1}}\right)}^2-2\frac{A_{3}}{A_{1}}\right] e^{-2\kappa_1 t}\left(r-r_1\right)^2+{\cal O}\left(\left(r-r_1\right)^3\right) \right\}~,
\end{equation}
and expanding both the static and the dynamic {terms} up to order ${\cal O}(r-r_1)$, one obtains that the diverging terms ($\propto(r-r_1)^{-2}$ and $\propto(r-r_1)^{-1}$) in $\rho_{\rm st}$ and $\rho_{{\rm dyn}}$ exactly cancel each other.\cite{Finazzi:2009jb}
%An analogous cancellation is found when studying the formation of a black hole through gravitational collapse~\cite{Barcelo:2007yk}. 
It is now clear that the total $\rho$ is ${\cal O}(1)$ on the horizon and does not diverge at any finite time. 
%(as expected from Fulling--Sweeny--Wald theorem, Ref.~\refcite{Fulling:1978ht}).
By looking at the subleading terms,
\begin{equation}
 \rho=\frac{e^{-2\kappa_1 t}}{48\pi}\left[3{\left(\frac{A_2}{A_1}\right)}^2-2\frac{A_3}{A_1}\right] + A
	+{\cal O}\left(r-r_1\right)~,
\end{equation}
where $A$ is a constant, we see that on the black horizon the contribution of the transient radiation (different from Hawking radiation) dies off {  exponentially with time, on} a time scale $\sim 1/\kappa_1$.
%\footnote{However, in analogy to the conclusions of~\cite{Barcelo:2007yk}, a slow approach to the black-horizon formation might lead to large values of the RSET and hence to a large back-reaction.}

Close to the white horizon, the divergences in the static and dynamical contributions cancel each other, as in the black horizon case. However, something distinctive occurs with the subleading contributions. In fact, they now becomes
\begin{equation}
 \rho=\frac{e^{2\kappa_2 t}}{48\pi}\left[3{\left(\frac{D_2}{D_1}\right)}^2-2\frac{D_3}{D_1}\right] + D
	+{\cal O}\left(r-r_1\right)~.
\end{equation}
This expression shows an exponential increase of the energy density with time. This means that {  $\rho$ grows exponentially and eventually diverges along \hm}.

In a completely analogous way, one can study $\rho$ close to {the Cauchy horizon}. 
Performing an expansion at late times ($t\to+\infty$) one finds that the RSET diverges also there.\cite{Finazzi:2009jb} 
%without any contradiction with the Fulling-Sweeny-Wald theorem, Ref.~\refcite{Fulling:1978ht}, because this is precisely a Cauchy horizon.
Note that the above mentioned divergences are very different in nature. The divergence at {late times on \hm}~stems from the untamed growth of the transient disturbances produced by the white horizon formation. The RSET divergence on {the Cauchy horizon} is due instead  to the well known infinite blue-shift suffered by light rays while approaching this kind of horizon. 
%It is analogous to the often claimed instability of inner horizons in Kerr-Newman black holes~\cite{Simpson:1973ua,poissonisrael,markovicpoisson}. 
While the second can be deemed inconclusive because of the Kay--Radikowski--Wald theorem, the first one is inescapable.
Summarising: the backreaction of the RSET will doom the warp drive making it semiclassically unstable.

\section{Extension to the Lorentz breaking case}

The just described semiclassical instability stems from standard, relativistic, QFT in curved spacetimes. 
One might wonder if the story could be different in scenarios where a UV completion of the theory is provided by some QG scenario.
This is the case of analogue gravity inspired Lorentz breaking scenarios (see e.g.~Ref.~\refcite{Barcelo:2005fc}) where generically one expects the standard relativistic dispersion relation for matter field to be replaced by $E^2=c^2 (p^2+ p^n/M_{\rm LIV}^{n-2})$ where $M_{\rm LIV}$ is normally assumed to be of the order of the Planck mass and $n$ is some integer greater than two.

Indeed, this is a modification that could potentially stabilise the warp drive, as it is by know understood that modified LIV dispersion relations are able to remove Cauchy horizons instabilities and tame the divergence of fluxes at white hole horizons. The reason for this is simple, UV rays in the above dispersion relations are faster or slower than light, in both cases light rays will not accumulate at the horizons (past or forward in time depending on the black or white nature of the horizon) as they normally do. Hence no built up of divergences can take place. 

Can this be a scenario where a quantum gravity inspired UV completion/regularisation could appear? 
This problem was dealt with in Ref.~\refcite{Coutant:2011fz} and surprising it leads to a negative answer, i.e.~not even the breakdown of Lorentz invariance can stabilise superluminal warp drives. Let us see how this works.

For the sake of simplicity we work in $1+1$ dimensions and consider a stationary situation. 
As in section \ref{sec:calculation}, we can define a new spatial coordinate
$X=x-v_c t$ (we use a different notation to avoid confusion between the two calculations) so the warp drive metric becomes 
%$ds^2=-c^2 dt^2+\left[dX-V(X)dt\right]^2$,
%
\begin{equation}\label{eq:pg}
 ds^2=-c^2 dt^2+\left[dX-V(X)dt\right]^2,
\end{equation}
where $V(X) = v_c(f(X) - 1)$ is negative. 
In this space-time, $\partial_t$ is a globally defined Killing 
vector field whose %R ok ?
 norm is given by $c^2 - V^2$: it is time-like within the bubble,
 its norm vanishes on the two horizons, and it is space-like outside. In a fluid flow analogy this would correspond to two superluminal asymptotic regions separated by a black and a white horizon from a compact internal subluminal region.\cite{Coutant:2011fz}
 
We can now consider a massless scalar field with a quartic dispersion relation. In covariant terms, its action reads 
\begin{equation}
  S_\pm=\frac{1}{2}\int\! d^2 x %\,\dd t\,
\sqrt{-g}\left[g^{\mu\nu}\partial_{\mu}\phi
\partial_{\nu}\phi \pm \frac{(h^{\mu\nu}\partial_{\mu}\partial_{\nu}\phi)^2}{M_{\rm LIV}^2}\right],
\label{action}
\end{equation}
where $h^{\mu\nu}=g^{\mu\nu}+u^\mu u^\nu$ is the 
spatial metric in the direction orthogonal to the unit time-like vector field $u^\mu$
which specifies the preferred frame used to implement the dispersion relation.
The sign $\pm$ in Eq.~(\ref{action}) holds for superluminal and subluminal dispersion, respectively. 

%In the present settings $u^\mu$ should be given from the outset, while in condensed matter the preferred frame is fixed by the fluid flow. Inspired by this analogy,  one can choose $u^\mu$ to be $(1,V)$ in the $t,X$ frame, {\it i.e.}\/ stationarity is preserved. Then the {\it aether} flow is geodesic  and it  is asymptotically at rest in the  $t,x$  frame of Eq.~(\ref{eq:3Dalcubierre}). 

Using Eq.~\eqref{eq:pg} and taking $u^\mu=(1,V)$ in the $t,X$ frame, the wave equation is 
\begin{equation}
 \left[\left(\partial_t+\partial_X V\right)\left(\partial_t+V\partial_X \right)-\partial_X^2\pm\frac{1}{M_{\rm LIV}^2}\partial_X^4\right]\phi=0. 
\label{modeequation}
\end{equation}
and $V(x)$ can be shaped so to mimic the warp drive geometry.
Because of stationarity, the field can be decomposed in stationary modes
$\phi = \int d\omega e^{- i \omega t }\phi_\omega$, where $\omega$ is 
the conserved (Killing) frequency. Correspondingly, at fixed $\omega$
the  dispersion relation reads 
\begin{equation}\label{eq:dispersion}
(\omega-V k_{\omega})^2=k_{\omega}^2\pm\frac{k_{\omega}^4}{M_{\rm LIV}^2}
\equiv\Omega_\pm^2,
\end{equation}
where $k_{\omega}(X)$ is the spatial wave vector, and $\Omega$ the {\it comoving} frequency, {\it i.e.}~the frequency in the aether frame.
The quartic nature of the dispersion relations allows up to four solutions/modes and the problem in the end reduces to solve a Bogoliubov matrix of coefficients relating the mode in the asymptotic regions (assuming $M_{\rm LIV}$ to be larger than any other scale in the problem).\cite{Coutant:2011fz}

The upshot is that in the case of subluminal dispersion relation there is an instability related to the well known ``laser effect".\cite{Corley:1998rk}
In the case of superluminal dispersion relations there is an infrared divergence that leads to a linear growth in time of the energy density proportional to $M_{\rm LIV}$ and the square of the warp drive wall surface gravity $\kappa$ (we are assuming $\kappa_1=\kappa_2=\kappa$).\cite{Coutant:2011fz}
Using quantum inequalities, Ref.~\refcite{Roman:2004xm}, one can argue that $\kappa$ must be of the order of the Planck scale,
which implies that the growth rate is also of that order (unless $M_{\rm LIV}$ is very different from that scale). 
So, even in the presence of superluminal dispersion, warp drives would be unstable on short time scales.

\section{Conclusions}

In summary, this investigations shows that the chronology protection conjecture could be implemented in a more subtle way than expected. Not only chronological horizons could be forbidden by quantum gravity, even the spacetime configuration that could be transformed into time machines could be unstable!
Noticeably, warp drives offer a first evidence in this sense, showing an instability which can be fully predicted within the realm of standard QFT in curved spacetime and not even theories with a preferred frame can escape this conclusion. 
It would be interesting if similar instabilities could be found in similar ``superluminal travel allowing spacetimes" such as traversable wormholes.

\section*{Acknowledgments}

The author wish to acknowledge that the research reported here was done in collaboration with C.~Barcle\`o, A.~Coutant, S.~Finazzi and R.~Parentani.
I also wish to acknowledge the John Templeton Foundation for the supporting grant \#51876.

\end{document}